\documentclass[aps,prl,twocolumn,showpacs,preprintnumbers,amsmath,amssymb,superscriptaddress]{revtex4-1}
\usepackage{graphicx}
\usepackage{bm}
\usepackage{braket}
\usepackage{natbib}

\begin{document}

\title{Measuring the Aharonov-Anandan phase in photonics}

\author{Kai Wang}
\email{Email: nick.kai.wang@gmail.com}
\author{Steffen Weimann}
\author{Stefan Nolte}
\author{Armando Perez-Leija}
\author{Alexander Szameit}
\email{Email: alexander.szameit@uni-jena.de }
\affiliation{Institute of Applied Physics, Abbe Center of Photonics, Friedrich-Schiller-Universit\"{a}t Jena, Max-Wien-Platz 1, 07743 Jena, Germany}

\date{\today}

\begin{abstract}
The Aharonov-Anandan phase is a description of the geometric nature in non-adiabatic cyclic evolutions of quantum states. Here we report on a measurement of the Aharonov-Anandan phase in photonics. We consider a time-independent quantum driven harmonic oscillator that is initially prepared at the vacuum state. We utilize evanescently coupled waveguides to realize this physical model and achieve a measurement of the Aharonov-Anandan phase via integrated interferometry.

\end{abstract}



\maketitle

Since its discovery by Berry in 1984~\cite{Berry1984}, the so-called geometric phase has been a subject of extensive study. In plain words, Berry's phase is an observable phase accumulation in eigenstates of physical systems governed by adiabatically changing Hamiltonians. In 1987, Aharonov and Anandan generalized this concept to the case of cyclic evolutions of states under the action of nonadiabatic Hamiltonians~\cite{Aharonov1987} by taking into account quantum mechanical systems that exhibit \emph{revivals} of the initial states, $\ket{\psi\left(T\right)}=\exp\left(i\phi\right)\ket{\psi\left(0\right)}$. By considering a projected state $\ket{\widetilde{\psi}\left(t\right)}=\exp\left(-iF\left(t\right)\right)\ket{\psi\left(t\right)}$ such that $\ket{\widetilde{\psi}\left(T\right)}=\ket{\widetilde{\psi}\left(0\right)}$ (and thus $\phi=F(T)-F(0)$), they find that the total phase $\phi$ consists of two parts, explicitly given by
\begin{equation}\label{eq:phase}
\begin{split}
\phi=-\int_{0}^{T}\bra{\psi\left(t\right)}\hat{H}\left(t\right)\ket{\psi\left(t\right)} \ \mathrm{d}t\\
+\int_{0}^{T}\bra{\widetilde{\psi}\left(t\right)}i\frac{\partial}{\partial t}\ket{\widetilde{\psi}\left(t\right)} \ \mathrm{d}t.
\end{split}
\end{equation}
The first and second terms on the r.h.s. of Eq.~\eqref{eq:phase} represent the dynamic and the Aharonov-Anandan phase (AA-phase), respectively.

The geometric phase provides a unified description for a variety of effects in physics~\cite{Aharonov1959,Thouless1982}. In addition to fundamental interests, the geometric phase and its non-Abelian generalizations can form the basis of any quantum computation~\cite{Zanardi1999,Ekert2000}. Since such phases are immune to certain types of errors~\cite{Filipp2009}, in particular random fluctuations during time evolution, they are potential tools for robust quantum computation and quantum information processing~\cite{Jones2000,Leibfried2003}.

Note that, in order to observe Berry's phase, the state of the system must remain stationary in such a way that there are no transitions to other instantaneous eigenstates. Such stationarity condition is guaranteed by the adiabatic theorem. On the contrary, for the case of nonadiabatic cyclic Hamiltonians, stationary states are only possible for time-independent systems. As a result, the total phase acquired by such stationary states after any period of time is purely dynamic and the AA-phase does not appear whatsoever. The situation is different when considering nonstationary states, the case in which the AA-phase becomes observable~\cite{Zeng1995}. This indicates that it is possible to observe AA-phases in cyclic time-independent systems. Another difference between Berry's and AA-phase is that in the former the geometric character is elucidated by mapping the states of the system into the space of the parameters of the Hamiltonian which are used to accomplish its time variations. In the AA-phase, the states are mapped into the projective space relative to the entire Hilbert space of the system.

Hitherto, the experiments of geometric phases were mostly performed in quantum systems, such as nuclear magnetic resonance (NMR)~\cite{Suter1988}, nuclear quadrupole resonance (NQR)~\cite{Tycko1987}, solid state qubit~\cite{Leek2007} and superconducting circuit~\cite{Pechal2012} systems. However, the measurements of geometric phases can also go beyond quantum systems to other platforms~\cite{Berry1984}, in which optical systems play an important role. The observation of Berry's phase was firstly achieved in optics using classical light in a fiber bundle~\cite{Tomita1986} and, later, in image-bearing beams~\cite{Segev1992}. A non-adiabatic-like topological phase was also observed by means of a non-planar Mach-Zehnder interferometer~\cite{Chiao1988}, yet its connection to the AA-phase is not direct and the origin of this phase can be explained by reflections~\cite{Jordan1988}. Thus looking for optical systems for measuring the AA-phase is still of great interest.

The quantum harmonic oscillator is among the few analytically solvable non-trivial archetypes of quantum physics. In fact, cyclic evolutions do exit in harmonic oscillator systems. Nevertheless, harmonic oscillators are usually difficult to implement in quantum systems, particularly in pure Fock states~\cite{Hofheinz2008}. Moreover, since the geometric phase is independent of the initial eigenstate, measuring the geometric phase in such harmonic systems is not achievable by simply comparing the phases between different eigenstates~\cite{Pechal2012,Leek2007}. To our knowledge, experimental investigations on the geometric phase of harmonic oscillators have been done in trapped ions~\cite{Leibfried2003} and superconducting circuits~\cite{Pechal2012}, not yet in photonics.

In this Letter we report on an observation of the AA-phase of the quantum driven harmonic oscillator (DHO) in integrated photonic lattices. Such systems form a highly-controllable platform to observe a variety of quantum effects, as they are well described by a discrete Schr\"odinger equation~\cite{Jones1965}. In our work we use integrated interferometry to observe the AA-phase acquired by light in a photonic realization of the DHO.

\begin{figure}
	\includegraphics{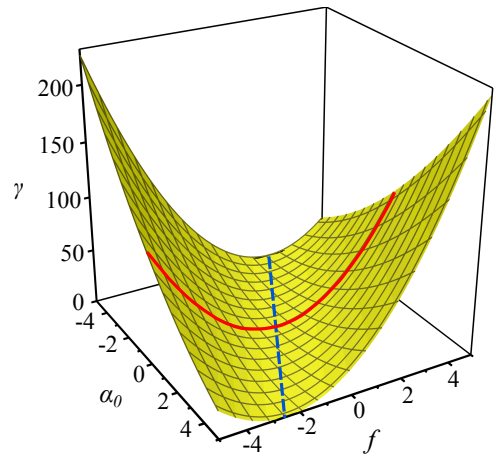}
	\caption{(Color online) Exemplary plot of the dependence of the AA-phase $\gamma$ on the coherent state parameter $\alpha_0 \in \mathbb{R}$ and the driving force $f$ for a DHO prepared in the coherent state $\ket{\alpha_0}$. The angular frequency $\omega=0.25$. The blue dashed curve corresponds to $\alpha_0=-\frac{f}{\omega}$ and the red solid curve represents the case $\alpha_0=0$. \label{para_3d}}
\end{figure}

To set the stage of this work, consider the Schr\"odinger equation for a DHO with time-independent parameters
\begin{equation}\label{eq:schr}
\left[\omega \left(\hat{a}^\dagger \hat{a}+\frac{1}{2}\right)+f(\hat{a}^\dagger +\hat{a})\right]\ket{\psi(t)}=i\frac{\partial}{\partial t}\ket{\psi(t)},
\end{equation}
where $\omega \in \mathbb{R}$ is the angular frequency of oscillation, and $f \in \mathbb{R}$ is essentially the driving force. In Eq.~\eqref{eq:schr} the raising operator $\hat{a}^\dagger$ and lowering operator $\hat{a}$ are defined for the quantum simple harmonic oscillator, and we take $\hbar=1$. By writing $\ket{\psi(t)}$ in the Fock space representation, $\ket{\psi(t)}=\sum_{n=0}^{\infty}c_n(t)\ket{n}$, and using the orthormality of the Fock states $\ket{n}$ and the properties of $\hat{a}$ and $\hat{a}^\dagger$, Eq.~\eqref{eq:schr} becomes a discrete Schr\"odinger equation
\begin{equation} \label{eq:GBL}
\begin{split}
i\frac{\mathrm{d}}{\mathrm{d} t} c_n(t)= & \omega \left(n+\frac{1}{2}\right) c_n(t) \\
&+ f\sqrt{n+1}c_{n+1}(t)+f\sqrt{n}c_{n-1}(t).
\end{split}
\end{equation}
The evolution operator of this Hamiltonian is given by
$\hat{U}(t)=e^{A(t)-i\frac{1}{2}\omega t}e^{-i\omega t \hat{a}^\dagger \hat{a}}e^{B(t)\hat{a}^\dagger}e^{-B^\ast(t)\hat{a}}$~\cite{Perez-Leija2012},
where $B(t)=-\frac{f}{\omega}\left(e^{i\omega t}-1\right)$ and $A(t)=\frac{f^2}{\omega^2}\left(e^{-i\omega t}+i\omega t-1\right)$.
Quite generally, if such a DHO is prepared in a coherent state $\ket{\psi(0)}=\ket{\alpha_0}$, the instantaneous wavefunction $\ket{\psi(t)}$ remains in a coherent state $\ket{\alpha(t)}$ just acquiring a time-dependent phase factor
\begin{equation}
\ket{\psi(t)}=e^{iF(t)} \ket{\alpha(t)},
\end{equation}
where the total phase $\phi(t)=F(t)-F(0)$ and $\alpha(t)$ is given by
\begin{equation} \label{beta}
\alpha (t)=\left(\frac{f}{\omega}+\alpha_0 \right)e^{-i\omega t}-\frac{f}{\omega}.
\end{equation}
We can see from Eq.~\eqref{beta} that, as long as $\ket{\alpha_0}$ is not an eigenstate of the Hamiltonian (i.e.  $\alpha_0\neq -\frac{f}{\omega}$), the evolution of the state is always periodic with a revival time of $T=\frac{2\pi}{\omega}$. Thus we can take $\ket{\alpha(t)}$ as the projected state $\ket{\widetilde{\psi}(t)}$ in Eq.~\eqref{eq:phase}.
Using the definition of the AA-phase~\cite{Aharonov1987}, we can show that at the revival time $(t=T)$, the AA-phase of the DHO is described by
\begin{equation}\label{projectedAA}
\gamma =  \int_0^T \! \alpha ^\ast(t) i\frac{\mathrm{d}}{\mathrm{d}t}\alpha(t)\,  \mathrm{d}t.
\end{equation}
According to Eq.~\eqref{projectedAA} and by using the Green's theorem, one finds that the AA-phase is -2 $\times$(the area covered by the path that $\alpha (t)$ traces out in the complex plane)~\cite{Chaturvedi1987}, which shows its geometric nature. In the particular case analyzed here, by inserting Eq.~\eqref{beta} into Eq.~\eqref{projectedAA} we obtain the AA-phase
\begin{equation}\label{phase_theo1}
\gamma=2\pi\left|\frac{f}{\omega}+\alpha_0\right|^2.
\end{equation}
In the case $\alpha_0 \in \mathbb{R}$, $\alpha (t)$ traces out a circle of radius $\frac{f}{\omega}+\alpha_0$ in the complex plain. Hence, Eq.~\eqref{phase_theo1} reduces to
\begin{equation}\label{phase_theo2}
\gamma=2\pi\left(\frac{f}{\omega}+\alpha_0\right)^2.
\end{equation}
For a given harmonic oscillator with constant $\omega$, in Fig.~\ref{para_3d} we plot the AA-phase $\gamma$ versus $f$ and $\alpha_0$ as an example. Note that $\gamma$ vanishes only for $\alpha_0=-\frac{f}{\omega}$ (see blue dashed curve in Fig.~\ref{para_3d}), which implies that the system was initially prepared at one eigenstate and thus it remains stationary over the cycle.
As a result, the total phase is purely dynamic.
However, in the region where $\alpha_0\neq -\frac{f}{\omega}$ (red solid curve in Fig.~\ref{para_3d}), the $\gamma$ is non-zero due to the non-adiabatic and nonstationary evolution of the state. Therefore, the AA-phase of such a time-independent system is indeed observable. If we choose the simplest case where the initial state of the DHO is the vacuum state~$\ket{0}$, i.e. $\alpha_0=0$, the AA-phase at the first revival is given by
\begin{equation} \label{AAphase}
\gamma=2\pi\left(\frac{f}{\omega}\right)^2.
\end{equation}
This equation reveals that the AA-phase depends quadratically on the driving force $f$.

\begin{figure}
	\includegraphics{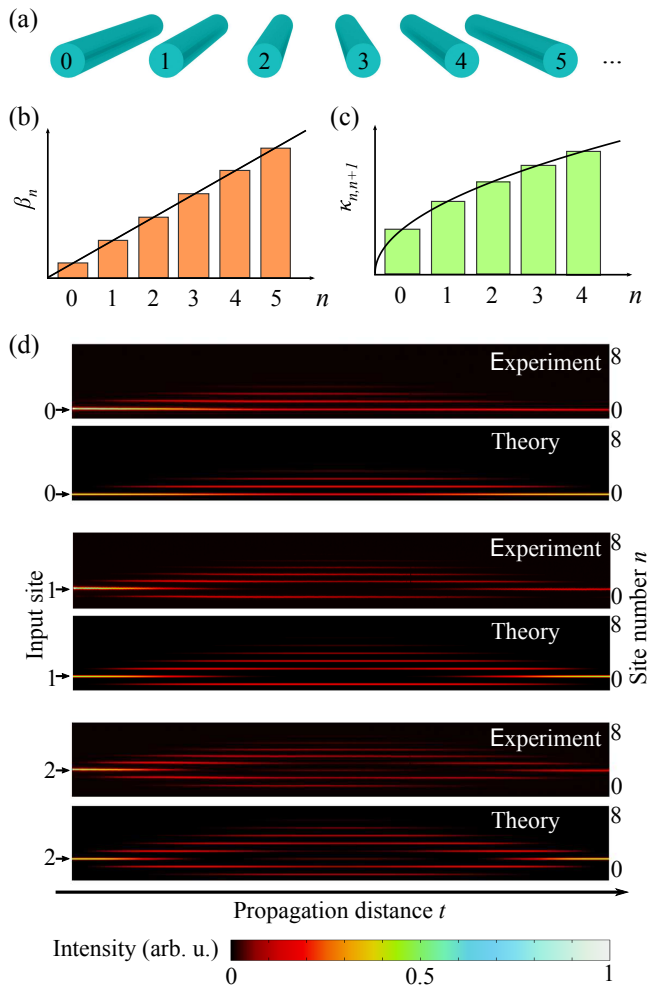}
	\caption{(Color online) (a)~Sketch of a Glauber--Bloch lattice. (b)~Linear dependence of the propagation constant $\beta_n$ of each waveguide on the site number $n$. (c)~The coupling constant $\kappa_{n,n+1}$ between adjacent waveguides $n$ and $n+1$ is proportional to $\sqrt{n+1}$. (d)~Experimental images from fluorescence microscopy showing the light evolution in a representative Glauber--Bloch lattice, compared with theory. The lattice consists of 9 waveguides. From top to bottom, light is launched into the same lattice from waveguide site 0, 1 and 2, respectively.
		\label{fig2}}
\end{figure}

In what follows, we show an experiment to measure such an AA-phase. Several investigations have demonstrated that it is possible to implement quantum systems in arrays of evanescently coupled waveguides~\cite{Longhi2009}. Here one can interpret Eq.~\eqref{eq:GBL} as a semi-infinite array of waveguides described by coupled mode theory~\cite{Jones1965}. In this optical context, $c_n(t)$ represents the light amplitude propagating in the $n$-th waveguide, $n$ ranging from $0$ to infinity (see Fig. \ref{fig2}(a)). Here $t$ is the propagation distance. The term $\omega \left(n+\frac{1}{2}\right)$ represents the propagation constant in a certain co-moving frame (see Fig. \ref{fig2}(b)). Moreover, the hopping rate $\kappa_{n,n+1}$ between adjacent waveguides $n$ and $n+1$ is proportional to $\sqrt{n+1}$, $\kappa_{n,n+1}=f\sqrt{n+1}$ (see Fig.~\ref{fig2}~(c)).
We term this optical arrangement as the Glauber--Bloch lattice, which is a photonic realization of the DHO in the Fock basis~\cite{Perez-Leija2012, Keil2012}. Analogous types of lattices have also been used for simulating the Rabi model~\cite{Crespi2012} and the Jaynes--Cummings model~\cite{Longhi2011}.

To experimentally realize the Glauber--Bloch lattice, we use the femto-second laser direct writing technique~\cite{Szameit2010}.
In the fabrication, femto-second pulses with a central wavelength of $515\ \mathrm{nm}$, a pulse duration of $305\ \mathrm{fs}$, a power of approximately $120\ \mathrm{mW}$ and a repetition rate of $200 \ \mathrm{kHz}$) are tightly focused into the bulk of a fused silica sample. The propagation constant of each waveguide is controlled by varying the velocity when we translate the sample~\cite{Szameit2010} through the fixed laser spot, and by changing the separation between adjacent waveguides we adjust the coupling constants.
In Fig. \ref{fig2}(d) we show experimentally the light evolution in a representative Glauber--Bloch lattice, which is observed by means of fluorescence microscopy~\cite{Szameit2007}. The lattice in Fig. \ref{fig2}(d) contains 9 waveguides with a revival length of approximately 83 mm with $f=0.03 \mathrm{mm^{-1}}$. The intensity of light in each waveguide $n$ represents the probability to populate the state $\ket{n}$. In Fig. \ref{fig2}(d) from top to bottom we show the excitation of site number 0, 1 and 2, respectively. In each case of excitation, we also show the theoretical prediction for comparison. We find that in all cases the intensity of light in each waveguide along propagation fits well the theoretical predictions. More specifically, after the excitation of the vacuum state $\ket{0}$, the intensity evolution at each $t$ is consistent with the corresponding probability distribution of the coherent state $\ket{\alpha(t)}$ in the Fock basis.

\begin{figure}
\includegraphics{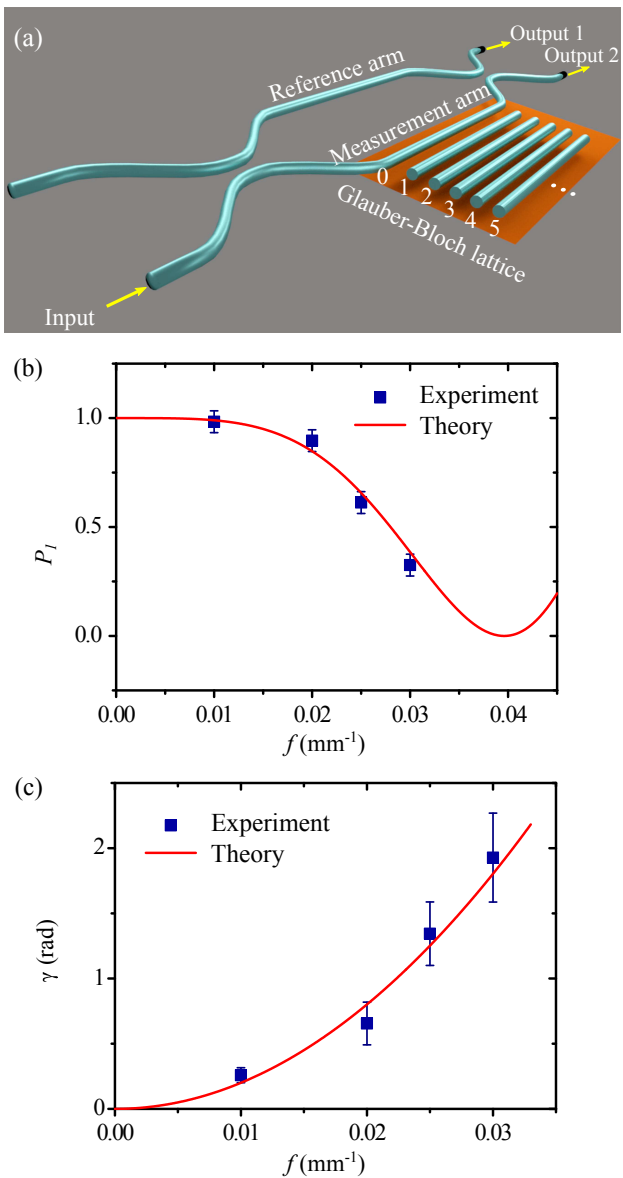}
\caption{(Color online) (a)~Schematic of the experimental set-up for measuring the AA-phase in a Glauber--Bloch lattice with an integrated MZI. (b)~Measured normalized output power of channel 1 $P_1$ as a function of the driving force $f$. (c)~The corresponding AA-phase deduced from the measured output powers given in (b).\label{fig3}}
\end{figure}

The total phase that a state accumulates at the revival consists of the dynamic phase and the AA-phase. A key point to measure the AA-phase lies in the removal of the dynamic phase. For a time-independent DHO, which is initially prepared in the vacuum state $\ket{0}$, it can be easily proven that the instantaneous dynamic phase is exactly the same as the total phase of the same harmonic oscillator without a driving force. In our experiment, such kind of simple harmonic oscillator can be implemented by a single waveguide with the same propagation constant as the $0$-th waveguide in the Glauber--Bloch lattice. Thus, by comparing the total phase of the single waveguide and the Glauber--Bloch lattice after a propagation distance of one revival we can measure the AA-phase.
The comparison is achieved by interferometry: As shown in Fig.~\ref{fig3}(a), we construct a Mach-Zehnder interferometer (MZI) using integrated waveguides. In the MZI, two directional couplers play the role of 50-50 beam splitters. The measurement arm of the MZI contains the lattice site $0$ of the Glauber--Bloch lattice, and the reference arm is a single waveguide aimed to cancel out the dynamic phase. The set-up shown corresponds to the case where the DHO is prepared at the vacuum state $\ket{0}$. The length of the Glauber--Bloch lattice corresponds to exactly one revival distance $T=\frac{2\pi}{\omega}$. At this specific distance, the state evolves back to the vacuum state $\ket{0}$ (see Fig.~\ref{fig2}(d) (top)) acquiring a total phase. As explained above, the dynamic phases acquired in both arms are equal. Thus at the second beam splitter of the MZI, the light fields of the two arms interfere, automatically eliminating the dynamic phase. Therefore, the power ratio of the two output ports will give direct information of the AA-phase:
\begin{equation}\label{eq:p1}
P_1=\cos ^2 \left(\frac{\gamma}{2}\right).
\end{equation}
Here, $P_1$ is the normalized power from output channel~$1$ (see Fig. \ref{fig3}(a)) of the MZI. Therefore, we can deduce the AA-phase from $P_1$ by
\begin{equation} \label{phase_p1}
\gamma=2\arccos\sqrt{P_1}.
\end{equation}

For our measurements of the AA-phase, we fabricate 4 Glauber--Bloch lattices each consisting of 5 waveguides with the same parameter $\omega=0.056\ \mathrm{mm^{-1}}$. The corresponding revival period in these lattices is $T=112 \ \mathrm{mm}$. In every Glauber--Bloch lattice, we choose a different driving force $f$. By doing so, we can measure and compare the AA-phases for the same harmonic oscillator with different driving forces.
As we launch light into one channel of the integrated MZI, we measure the powers at the output ports to obtain $P_1$. The points in Fig. \ref{fig3}(b) are the measured $P_1$ for different values of $f$. From Eq.~\eqref{AAphase} and \eqref{eq:p1} we see that $P_1$ depends on $f$ according to
\begin{equation}
P_1=\cos ^2 \left(\frac{\pi}{\omega^2}f^2\right),
\end{equation}
which is indicated by the red solid curve in Fig. \ref{fig3}(b). We find the measured $P_1$ to be consistent with our theoretical predictions.
By using Eq.~\eqref{phase_p1} we calculate the values of AA-phase for the lattices from the obtained $P_1$ (see the points in Fig. \ref{fig3}(c)). The theoretical prediction is depicted by the red solid curve according to Eq.~\eqref{AAphase}. We find the deduced AA-phase is in agreement with the theory.

In conclusion, we have performed a measurement of the Aharonov-Anandan phase in photonics. We utilize a waveguide-based platform to realize the Hamiltonian of quantum driven harmonic oscillators in Fock space and measure the phases using integrated Mach-Zehnder interferometers. We consider the case where the Hamiltonian of the DHO is time-independent and prepare the system initially at the vacuum state, which is not an eigenstate of the Hamiltonian. Thereby we obtain nonstationary states that evolve cyclically. We experimentally demonstrate the non-zero Aharonov-Anandan phase that depends quadratically on the driving force. The experimental platform is highly controllable and can be extended to further investigations of geometric phases for more complicated cases, such as other initial states and time-dependent Hamiltonians. Similar types of set-ups may also be used for potential applications of geometric phases in topological photonics~\cite{Lu2014,Rechtsman2013,Hafezi2013}.

The authors gratefully acknowledge financial support from the German Ministry of Education and Research (Center for Innovation Competence programme, grant 03Z1HN31) and the Deutsche Forschungsgemeinschaft (grants NO462/6-1 and SZ276/7-1).

\end{document}